%% file: main.tex
\documentclass[10pt]{article}

\input{preamble.tex}

\usepackage{ass_macros}

\usepackage{lmodern}
\usepackage{amssymb,amsfonts,amsmath,mathtext,enumerate,float}
\usepackage[hyphens,spaces,obeyspaces]{url}
\usepackage{hyperref}
\usepackage{wrapfig}
\usepackage{amsmath}
\usepackage{subfigure}

\usepackage{epsfig}
\usepackage{indentfirst}
\usepackage[usenames]{color}
\usepackage{tikz}

\usepackage[authoryear, round, numbers, comma, sort]{natbib}
\hypersetup{
    colorlinks=true,
    linkcolor=blue,
    filecolor=blue,      
    urlcolor=blue,
    citecolor=blue,
}

\def\ursa{{XTE\;J1118$+$480}}
\def\mon{{A0620$-$00}}
\def\musc{{GRS\;1124$-$68}}

\begin{document}
\pagestyle{headings}
\title{
\rule{\textwidth}{1pt}
\\ 
Change in the orbital period of a binary system due
to an outburst in a windy accretion disc

\vspace{-0.3cm}
\rule{\textwidth}{1pt}
}

\authorname{A. L. Avakyan$^{1,2}$\thanks{\ \ E-mail: cygnusxonexray@gmail.com}, G. V. Lipunova,$^{1}$, K. L. Malanchev$^{1,3}$, N.I. Shakura$^{1,4}$}

\authoraddr{$^{1}$Lomonosov Moscow State University, Sternberg Astronomical Institute, Moscow, Russia \\
$^{2}$Lomonosov Moscow State University, Faculty of Physics, Moscow, Russia
\\
$^{3}$University of Illinois at Urban--Champaign, Urbana, Illinois, USA
}

\secondauthoraddr{$^{4}$Kazan Federal University,  Kazan, Russia}

\maketitle

\abstract
We consider a new mechanism for the removal of the angular momentum from an X-ray binary system and the change in its orbital period --- the mass loss in the form of a wind from an accretion disc. Both observations and models predict powerful winds from discs in X-ray transients. We have obtained an analytical estimate for the increase in the orbital period of a binary system with a wind from the disc during an outburst,  and quantitative estimates are given for the systems \ursa{}, \mon{} and \musc{}. Resulting rate of period grow  is of order of the observed rates of secular decrease in the period. We also compare the predicted rates of change in the period of a binary system due to the flow of matter into the disc and outflow from the second Lagrange point with the observations. It is concluded that the above mechanisms cannot explain the observed secular decrease in the period of the three X-ray Novae, and it is necessary to consider a circumbinary disc that drains the binary's angular momentum.


\section{Introduction}
Orbital period variations provide important information about the parameters of a close binary system, as well as the processes taking place inside it: the flow of matter between the components, the evolution of a companion star, etc. In close binaries, the change in the orbital period is mainly associated with a decrease in the angular momentum~\citep{Tut_Yun71, cherep}. This decrease in the angular momentum of a close binary system can be caused by various mechanisms, among which there are three principal ones: mass loss by the system~\citep{Rappaport82}, magnetic braking~\citep{Verbunt81, Tut83}, and gravitational waves~\citep{Landafshitz}.

In particular, the period changes are observed in low-mass X-ray binaries~(LMXB). In these systems, matter flows from the optical star to the compact component, accompanied by the formation of a hot accretion disc, which produces X-ray radiation.

\citet{Gonzalez12} (hereinafter GH12) were able first to detect a change in a LMXB orbital period when analyzing the observations of the~\ursa{} system. {It turned out that the orbital period decreased in the system.} In later works~\citep[][hereinafter GH14 and GH17]{Gonzalez14, Gonzalez17}, a decrease in the period was also found for \mon{} and \musc{}. The authors believe that the main cause of the rapid decrease in the period in these LMXB is magnetic braking; however, the model values that they have obtained, even under an assumption of a sufficiently strong magnetic field, are one or two orders of magnitude lower than the observed ones. 

In this paper, we consider another mechanism for the loss of angular momentum from a binary system --- the mass loss by the system. 
LMXBs exhibit repeated outbursts caused by instabilities in the disc or unsteady flow of matter between the components~\citep[e.g.,][]{DIM, Bath_etal74, Bath75, Bath_Pringle, Hameury}. During an outburst, the accretion rate on the compact component rises by {several orders of magnitude}, and 
most radiation from a LMXB goes in the X-ray range~\citep[see, for example][]{Chen97}. Presumably, during an outburst there is an outflow in the form of a wind from the accretion disc around the compact object. 

The presence of such a wind in the LMXBs is supported by modern observations indicating the expansion of ionized matter. Outflows from accretion discs are observed both as X-ray absorption lines shifted to the blue part of the spectrum~\citep[e.g.,][]{Ueda, Trigo}, and  as line broadening in the optical range~\citep[e.g.,][]{Munoz, Casares, Charles}. 

In most cases, X-ray absorption lines of highly ionized Fe are observed in systems with the inclination of more than $50^{\circ}$.
Therefore, the absorbing plasma should have a higher density closer to the disc, which suggests that it is a substance flowing out of the disc~\citep{HigProg}.

Observational rates of the mass loss in the wind are obtained for the intermediate mass X-ray binary~(IMXB) Her~X-1 in the work of~\citet{Kosec} and  the estimates strongly depend on the wind geometry. According to the paper, if the wind moves along the disc then the mass loss rate in the wind in Her~X-1 is approximately equal to the accretion rate onto the compact object. However, if the wind is spherically symmetric, then the mass-loss estimate increases by an order of magnitude. In the work by~\citet{Ponti12},  observational estimates of the rate of matter outflow from several X-ray transients are presented, from which it follows that the ratios of the rate of matter loss in the wind to the accretion rate onto a black hole are in the range from $1$ to $10$. In hydrodynamical numerical simulations of thermally driven disc winds~\citep{Higgin17, Luketic, Higgin19}, this ratio is obtained to lie in the range from $2$ to $15$. These studies presume that the heating of the matter occurs due to the irradiation of the outer parts of a disc by a central X-ray source. Thus, both modelings and observations favour strong winds of matter from the discs.

The problem of a change in the orbital period of a binary system as a result of the mass loss by a component (or both) has been studied earlier. The simplest description involves an isotropic gas release without taking into account the effect of an ejected matter on the motion of the binary system, and it was solved by~\citet{jeans}. It was assumed that the ejected matter was removed very quickly without affecting the orbital motion. This can be applicable in the case of strong outbursts of novae and supernovae. If the ejection velocity is not high enough, then the ejected matter affects the orbital motion of the binary system, not only due to the changes in its mass and angular momentum, but also due to the gravitational effect on the system. In addition, the ejected matter can return to the system to any of the stars, causing additional changes in the period. For example, a viscous toroidal disc around a binary system (circumbinary disc) may be formed, which can effectively transport the angular momentum it receives due to tidal interactions~\citep[][hereinafter CP19]{ChenPod}.

In addition to the wind from the accretion disc, outflow of matter from the outer Lagrange point $ L_2 $, which is located behind the less massive component ~ (behind the optical star in the case of LMXB), can occur in the system. And, unlike the wind from the disc, which is most active only during an outburst, the flow of matter from $ L_2 $ can be constantly present and lead to a constant decrease in the period of the system.

The article by~\citet{Changing} provides a detailed solution to the binary system mass-loss problem. In its formulation, it is necessary to know the components of the velocity and the coordinates of the ejected substance, which are not known with a sufficient accuracy. In the present paper, a simpler problem is considered that does not take into account the effect of the ejected matter on the orbital motion of the system as it goes to infinity.

The structure of the paper is as follows. In Section \S~\ref{model} the  equations of the orbital period change model are described in detail. Namely, subsections \S~\ref{model_wind}, \S~\ref{model_inflow}, \S~\ref{model_L2} are devoted to three different mechanisms: disc wind, inflow of matter form the donor star, and outflow from Lagrange point $L_2$, respectively. In section \S~\ref{results} we present the results of our model and the application to \ursa{}, \mon{}, and \musc.  The discussions and conclusions are provided in sections \S~\ref{discus} in \S~\ref{conclus}.

\section{Model of  orbital period change}\label{model}
\subsection{Impact of accretion disc wind}\label{model_wind}

Let the matter of the companion star flow through the Lagrange point $ L_1 $ into the Roche lobe of the compact object and become part of the accretion disc there. Under the influence of self-irradiation, magnetic field or radiation pressure, some of the matter can be carried away from the disc in the form of a wind, taking with it both the mass and the angular momentum. Thus, the total mass and total angular momentum of the binary system change. We will assume that all the wind starts from a fixed radius in the disc.

Let's calculate the change in the orbital period of the system. We will use the Kepler's third law:
    \begin{equation}\label{kep}
    P =  2\pi \, \sqrt{\frac{A^3}{GM}},
    \end{equation}
where $ M $ is the total mass in the binary system, and $ P $ and $ A $ are the period and semi-major axis, respectively.
Next, we find the relative change in the orbital period of the binary $ \Delta P / P $,
by varying~\eqref{kep} as a function of $ A $ and $ M $. As a result, we get:
    \begin{equation}\label{per}
    \frac{\Delta P}{P} =  \frac{3}{2}\frac{\Delta A}{A} - \frac{1}{2}\frac{\Delta M}{M},
    \end{equation}
where $ P $ and $ \Delta P $ are the orbital period of the system and its change, respectively. Our ultimate goal is to directly relate $ \Delta M / M $ and $ \Delta P / P $. Therefore, it is necessary to express $ \Delta A / A $ through $ \Delta M / M $.
 
To calculate this change in the semi-major axis, we use the law of conservation of angular momentum. We assume that before an outburst the disc is a ring of mass $ M_{\rm disc} $ with a characteristic radius $ R_{\rm disc} $. Further, all the 'primed' variables (when considering the effect of the wind) 
denote the state of the system after the outburst, and all the 'unprimed' ones, the state before the outburst.

\begin{figure}[ht!]
    \begin{center}
    \includegraphics[width=\columnwidth]{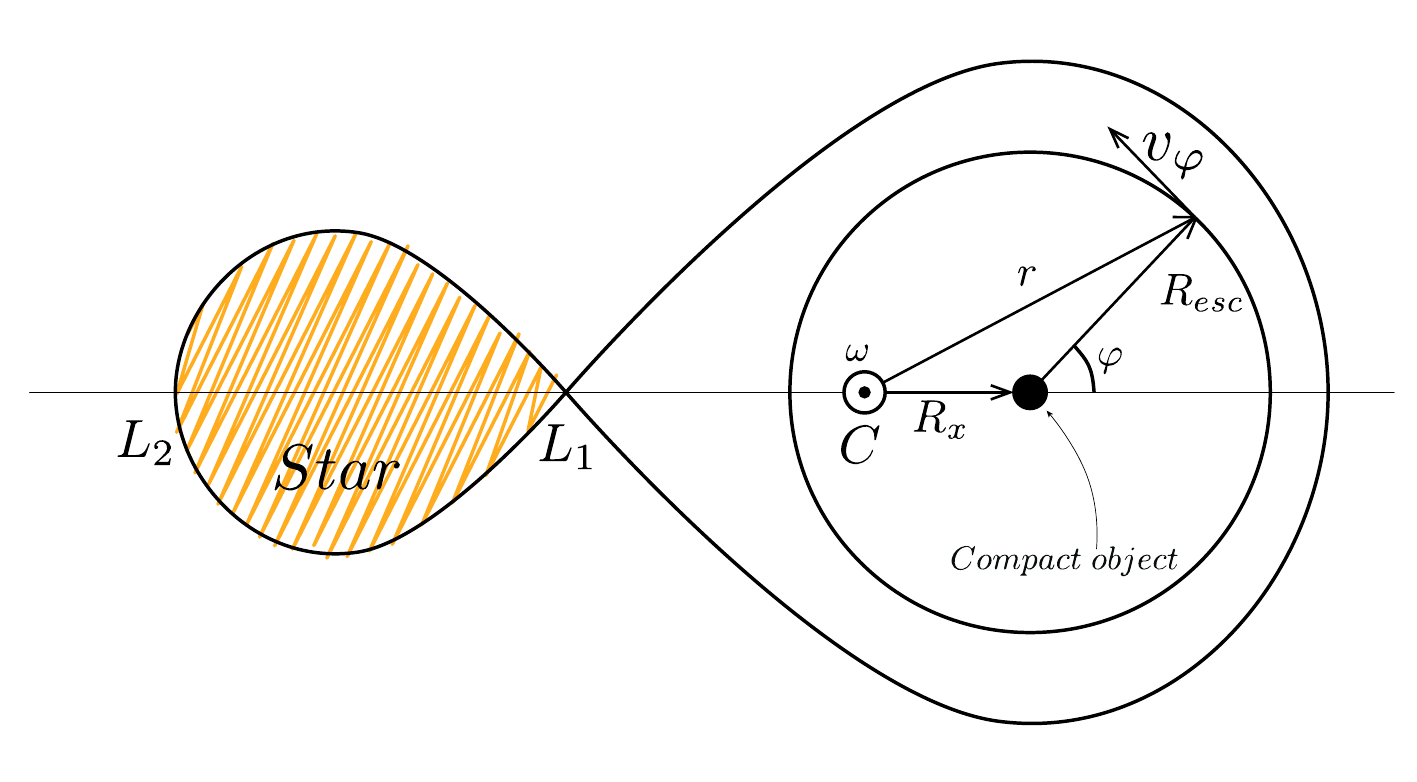}
    \caption{\rm Designations for a close binary system. Point C is the center of mass of the system, $ L_{1} $ and $ L_2 $ are the Lagrange points.}
    \label{system}
    \end{center}
    \end{figure}

The total angular momentum of the binary system before the ejection of matter to infinity $ J $ is:
    \begin{equation}\label{J}
    J = J_{\rm opt}  + J_{\rm x} + J_{\rm disc} ,
    \end{equation}
where $ J_{\rm opt} $, $ J_{\rm x} $, and $ J_{\rm disc} $ are the angular momentum of the optical star, compact object, and accretion disc around it, respectively. They are expressed
as follows:
    \begin{equation}\label{J_all}
    \begin{gathered}
    J_{\rm opt}  = M_{\rm opt} \, \omega_{\rm orb}  \, R_{\rm opt} ^2,\\
    J_{\rm x}  = M_{\rm x} \, \omega_{\rm orb}  \, R_{\rm x} ^2,\\
    J_{\rm disc}  = M_{\rm disc}  (\omega_{\rm orb} \, R_{\rm x} ^2 + \sqrt{G M_{\rm x}  R_{\rm disc} }).
    \end{gathered}
    \end{equation}
It can be seen that the angular momentum of the disc consists of the orbital angular momentum and the angular momentum associated with the Keplerian rotation around the relativistic star. $ M_{\rm opt} $ and $ M_{\rm x} $ are the masses of the optical and compact components, $ R_{\rm opt} $ and $ R_{\rm x} $ are the distances from centers of mass of stars to the center of mass of the binary system, $ \omega_{\rm orb} $ is the angular velocity of orbital rotation~(see Fig.~\ref{system}):
    \begin{equation}\label{J_par}
    \begin{gathered}
    R_{\rm opt}  = A\,(M_{\rm x} +M_{\rm disc} )/M,\\
    R_{\rm x}  = A\, M_{\rm opt}  / M,\\
    \omega_{\rm orb}  = \sqrt{\frac{GM}{A^3}} \, ,\\
    M = M_{\rm opt} + M_{\rm x}  + M_{\rm disc} \,.
    \end{gathered}
    \end{equation}
We assume that after the outburst the accretion disc is completely consumed: part of the disc mass,  $ \Delta M_{\rm acc}$, has fallen onto the compact object, as a result of which its mass increased to $M_{\rm x}^{'} = M_{\rm x} + \Delta M_{\rm acc}$, and the rest of the disc mass, $\left|\Delta M_{\rm wind}\right| = M_{\rm disc} - \Delta M_{\rm acc}$~($\Delta M_{\rm wind} < 0$), has flown out of the system as a wind. Thus, the angular momentum of the system after the outburst:
    \begin{equation}\label{J_2}
    \begin{gathered}
    J^{'} = J_{\rm opt} ^{'} + J_{\rm x} ^{'} \, , \\
    J_{\rm opt}^{'}  = M_{\rm opt} \, \omega_{\rm orb}^{'}  \, (R_{\rm opt}^{'}) ^2\,,\\
    J_{\rm x}^{'}  = (M_{\rm x} + \Delta M_{\rm acc}) \, \omega_{\rm orb}^{'}  \, (R_{\rm x} ^{'})^2\ + \Delta M_{\rm acc}\sqrt{GM_{\rm x}R_{\rm in}},
    \end{gathered}
    \end{equation}
where $ R_{\rm in} $ is the inner radius of the disc~(the radius of the last stable orbit).
We neglect the angular momentum $\Delta M_{\rm acc}\sqrt{GM_{\rm x}R_{\rm in}}$ added to the angular momentum of a compact object as a result of accretion, since the inner radius of the disc is $ R_{\rm in} $ is much smaller than other characteristic radii of the problem.
After the outburst, the quantities $ R_{\rm opt} $, $ R_{\rm x} $, $ w_{\rm orb} $, and $ M $ acquire new values, namely:
    \begin{equation}\label{J_2_par}
    \begin{gathered}
    R_{\rm opt}^{'}   = A^{'}\,(M_{\rm x} +\Delta M_{\rm acc} )/M^{'} \, ,\\
    R_{\rm x}^{'}  = A^{'}\, M_{\rm opt}  / M^{'} \,,\\
    \omega_{\rm orb}^{'}  = \sqrt{\frac{GM^{'} }{(A^{'})^3}} \\
    M ^{'} = M_{\rm opt} + M_{\rm x}  + \Delta M_{\rm acc} \,,
    \end{gathered}
    \end{equation}
where $ A^{'} $ is the value of the semi-major axis after the outburst. The change in the angular momentum of the $ J^{'} - J $ system is equal to the angular momentum carried away in the wind $ \Delta J_{\rm wind} $ ($ \Delta J_{\rm wind} < 0 $). If the wind is launched from the radius $ R_{\rm esc} $, then it carries the angular momentum of a ring with mass $ \left|\Delta M_{\rm wind} \right| $ participating in the orbital motion and rotation around the compact object:
    \begin{equation}\label{mod}
    \Delta J_{\rm wind} = 
    \Delta M_{\rm wind}(\omega_{\rm orb}  R_{x}^2 + \sqrt{GM_{\rm x}A}\sqrt{k_{\rm esc}}) < 0\,,
    \end{equation}
where $k_{\rm esc} \equiv R_{\rm esc}/A $. 
Here we take into account the possibility that the characteristic radius of the disc changes during an outburst: the outflow of the wind can occur from a radius larger than the characteristic radius of the disc in a quiescent state~($ R_{\rm disc} $).

By substituting~\eqref{mod} into the equation~$J = J^{'} - \Delta J_{\rm wind}$~($\Delta J_{\rm wind} < 0$), we find the 
relation between the semi-major axis change due to the wind influence $\Delta A_{\rm burst} \equiv A^{'} - A$ with the decrease in the total mass of the system and the accreted mass:
    \begin{equation}\label{a/a1_update}
    \begin{gathered}
    \frac{1}{2}\frac{\Delta A_{\rm burst}}{A} = \left(\frac{(M_{\rm opt} + M_{\rm x})^{3/2}}{M_{\rm opt}\sqrt M_{\rm x}}\,(\sqrt{k_{\rm esc}} - \sqrt{k_{\rm disc}}) - \frac12\right) \times \\
    \times \,\frac{\Delta M_{\rm wind}}{M_{\rm opt} + M_{\rm x}}\, + \, 
    \frac{(M_{\rm opt} + M_{\rm x})^{3/2}}{M_{\rm opt}\sqrt M_{\rm x}}\,\sqrt{k_{\rm disc}}\,\frac{\Delta M_{\rm acc}}{M_{\rm opt} + M_{\rm x}}\,,
    \end{gathered}
    \end{equation}
where $k_{\rm disc} \equiv R_{\rm disc}/A$ by analogy with $k_{\rm esc}$. 

If we divide  expression \eqref{a/a1_update} by the outburst duration, we can express the average rate of change of the semi-major axis in terms of the average accretion rate and the average rate of mass loss to the wind. Then, according to~\eqref{per}, the period change in terms of $  \left<\dot{P}_{\rm burst}\right>$, $\left<\dot{M}_{\rm wind}\right>$, $\left<\dot{M}_{\rm acc}\right>$ and $q=M_{\rm x} / M_{\rm opt} $ is written as follows:
    \begin{equation}\label{pwind/p_update}
    \begin{gathered}
    \frac{\left<\dot{P}_{\rm burst}\right>}{P} = \frac{3}{1 + q}\left( \frac{(1 +q)^{3/2}}{\sqrt {q}}(\sqrt{k_{\rm esc}} - \sqrt{k_{\rm disc}}) - \frac{2}{3} \right) \times \\
    \times \, \frac{\left<\dot{M}_{\rm wind}\right>}{M_{\rm opt}} \,+ \, \frac{3\,\sqrt{1+q}}{\sqrt{q}}  \sqrt{k_{\rm disc}} \;\frac{\left<\dot{M}_{\rm acc}\right>}{M_{\rm opt}}\, .
    \end{gathered}
    \end{equation}
Since $\left<\dot{M}_{\rm wind}\right> < 0 $, the first term in  equation~\eqref{pwind/p_update} is negative under the condition $k_{\rm esc}^{1/2} > k_{\rm disc}^{1/2} + 2/3 \, (q^{1/2}/( 1 + q )^{3/2})$. The second term is positive and is due to the accretion onto the compact object.

It is also necessary to take into account the following limitation following from the physics of the process. As mentioned above, the angular momentum of the disc $ J_{\rm disc} $ consists of the orbital angular momentum $ J_{\rm disc}^{\rm orb} $ and the angular momentum associated with the Keplerian rotation around the compact object $ J_{\rm disc}^{\rm Kepl} $:
    \begin{equation}\label{kepl_J}
    \begin{gathered}
    J_{\rm disc} = J_{\rm disc}^{\rm orb} + J_{\rm disc}^{\rm Kepl} = \\ =   M_{\rm disc} \, \omega_{\rm orb} \, R_{\rm x} ^2 + M_{\rm disc} \sqrt{G M_{\rm x}  R_{\rm disc} }.
    \end{gathered}
    \end{equation}
The angular momentum of the disc, associated with the rotation of the matter around the black hole, can only decrease during an outburst:  the disc's mass decreases and the disc is decelerated by the gravitational tidal forces near its outer boundary. As a result, the second term $J_{\rm disc}^{\rm Kepl}$  after the burst is divided between three non-negative parts:
(1) angular momentum  $J_{\rm acc}^{\rm Kepl}$ of the matter falling on a compact object;
(2) angular momentum $J_{\rm wind}^{\rm Kepl}$, carried away by the wind and
(3) angular momentum $J_{P}^{\rm Kepl}$, converted into orbital motion due to tidal forces near the outer radius of the disc:
    \begin{equation}\label{kepl_J_1}
    \begin{gathered}
    J_{\rm disc}^{\rm Kepl}  = J_{\rm acc}^{\rm Kepl} + J_{\rm wind}^{\rm Kepl}+ J_{P}^{\rm Kepl}\,, \\
    M_{\rm disc} \sqrt{G M_{\rm x}  R_{\rm disc} } =  \Delta M_{\rm acc} \sqrt{G M_{\rm x}  R_{\rm in}} \, + \\ + \, \left|\Delta M_{\rm wind}\right| \sqrt{G M_{\rm x}  R_{\rm esc}}  + J_{P}^{\rm Kepl}\,,
    \end{gathered}
    \end{equation}
As before, the part of the angular momentum added to the angular momentum of the compact object as a result of accretion $J_{\rm acc}^{\rm Kepl}$ is neglected because it is small compared to $J_{\rm wind}^{\rm Kepl}$ and $J_{\rm disc}^{\rm Kepl}$~($R_{\rm in} \ll  R_{\rm esc}$).

Accordingly, it follows from \eqref{kepl_J_1}  that  $\left|\Delta M_{\rm wind}\right| \sqrt{G M_{\rm x}  R_{\rm esc}}  \leq  M_{\rm disc} \sqrt{G M_{\rm x}  R_{\rm disc} }$.
Taking this into account and introducing the parameter $C_{\rm w} \equiv \left|\Delta M_{\rm wind}\right|/ \Delta M_{\rm acc}$, we get the restriction:
    \begin{equation}\label{C_wmax}
    \begin{gathered}
    C_{\rm w} \leq C_{\rm w}^{*} \equiv \frac{\sqrt{k_{\rm disc}}}{\sqrt{k_{\rm esc}} - \sqrt{k_{\rm disc}}}\, .
    \end{gathered}
    \end{equation}
It follows that for $C_{\rm w} = C_{\rm w}^{*}(q)$, that is, when the entire initial Keplerian angular momentum of the disc $ J_{\rm disc}^{\rm Kepl} $ is carried away by the wind, the change in the orbital period of the binary is described by the following formula~(Jeans model):
    \begin{equation}\label{macc_0_}
    \begin{gathered}
    \frac{\left<\dot{P}_{\rm burst}\right>}{P} = - \frac{2}{1 + q} \, \frac{\left<\dot{M}_{\rm wind}\right>}{M_{\rm opt}}.
    \end{gathered}
    \end{equation}
In particular,  formula~\eqref{macc_0_}  is applicable when all the ring's matter leaves the system without an accretion event. But such an approximation ~ (wind without accretion) for an X-ray nova is unjustified.

Note that the value of $C_{\rm w}^{*}(q)$ imposes a restriction on the parameter $C_{\rm w}(q)$, that is, on the wind power, only if the launch radius is greater than the initial radius of the disc, i.e. $k_{\rm esc} > k_{\rm disc}$ (in this case formula ~\eqref{C_wmax} makes sense). In other cases, the relative mass loss in the wind is not limited.

Using the introduced parameter $C_{\rm w}$,, we finally rewrite the formula~\eqref{pwind/p_update} as:
    \begin{equation}\label{pb_p}
    \begin{gathered}
    \frac{\left<\dot{P}_{\rm burst}\right>}{P} = 3\left[\,\sqrt{\frac{1 +q}{q}} \sqrt{k_{\rm disc}}\,(C_{\rm w}+1) + \frac{2}{3}\,\frac{C_{\rm w}}{1+q} \right. - \\ 
    - \left. \sqrt\frac{1+q}{q}  \sqrt{k_{\rm esc}}\, C_{\rm w} \right]\frac{\left<\dot{M}_{\rm acc}\right>}{M_{\rm opt}}\, .
    \end{gathered}
    \end{equation}
    
It should be noted that, taking into account the obtained limitation~\eqref{C_wmax} on the ratio of the mass loss rate  in the wind to the accretion rate, the total effect always leads to an increase in the period.

\subsection{Flow of matter into accretion disc}\label{model_inflow}
\noindent

Now let us turn to another source of the change in the period, namely, to the flow of matter from the optical component of the system into the Roche lobe of a compact object. In this case, it is also necessary to use the formula~\eqref {per}, but  now under the assumption that the parcel of mass $ \Delta M_{\rm tr} $~($\Delta M_ {\rm tr} > 0 $) is transferred from the star with point mass $ M_{\rm opt} $ and arrives into the accretion disc at the radius $ R_{\rm disc} $ without changing the total mass of the system ($ \Delta M = 0 $):
    \begin{equation}\label{ptr_var}
    \frac{\Delta P_{\rm tr}}{P} = \frac{3}{2}\frac{\Delta A_{\rm tr}}{A}\,,
    \end{equation}
where $\Delta P_{\rm tr}$ and $\Delta A_{\rm tr}$ are the  changes in the orbital period and semi-major axis, respectively.
We also write out the angular momentum before and after the matter transfer from the optical component into the accretion disc:
    \begin{equation}\label{J_tr}
    J = M_{\rm x}M_{\rm opt}\sqrt{\frac{GA}{M_{\rm x} + M_{\rm opt}}}\,,
    \end{equation}
    \begin{equation}\label{J_2_tr}
    \begin{gathered}
    J^ {''} =  M_{\rm x} \, \omega_{\rm orb}^{''}  \, (R_{\rm x}^{''}) ^2 + (M_{\rm opt} - \Delta M_{\rm tr})\, \omega_{\rm orb}^{''}  \, (R_{\rm opt}^{''}) ^2 \, + 
    \\
    + \, \Delta M_{\rm tr}  (\omega_{\rm orb}^{''} \, (R_{\rm x} ^{''}) ^2 + \sqrt{G M_{\rm x}  R_{\rm disc} }) \,,  
    \end{gathered}
    \end{equation}
where 
    \begin{equation}\label{J_all_tr}
    \begin{gathered}
    R_{\rm x}^{''}  = A^{''}\, (M_{\rm opt} - \Delta M_{\rm tr}) / M ,\\
    R_{\rm opt}^{''}  = A^{''}\,(M_{\rm x} + \Delta M_{\rm tr} )/M ,\\
    \omega_{\rm orb}^{''}  = \sqrt{\frac{GM}{(A^{''})^3}}\, , \\
    A^{''} = A + \Delta A_{\rm tr}\, ,\\
    M = M_{\rm x} + M_{\rm opt} \,.
    \end{gathered}
    \end{equation}
In this section, by analogy with the previous one, the 'double-primed' variables denote the state of the system after the matter transfer, and the 'unprimed' subscript indicates the state before that.

Taking into account that  $J = J^{''}$ (the total orbital angular momentum does not change), as well as the expression~\eqref{ptr_var}, we obtain a formula for changing the period of the binary system due to the flow of matter into the disc around the compact object. Expressing everything in terms of $\left<\dot{P}_{\rm tr}\right>$, $q$, and $\left<\dot{M}_{\rm tr}\right>$:
    \begin{equation}\label{ptr/p_update}
    \frac{\left<\dot{P}_{\rm tr}\right>}{P} = \frac{3}{(1+q)}\left(  q -  \frac1q- \frac{(1+q)^{3/2}}{\sqrt{q}}\sqrt{k_{\rm disc}} \right)\frac{\left<\dot{M}_{\rm tr}\right>}{M_{\rm opt}}.
    \end{equation}
The above is reduced to the classical formula describing the flow of matter between the point masses~\citep[see, for example,][]{cherep} by zeroing $k_{\rm disc}$.

Since we consider low-mass X-ray binaries, in which a star of a lower mass is the donor, the rate of the change in the period due to the flow of matter into the disc is always positive~($\left<\dot{P}_{\rm tr}\right>/P > 0$) and leads to an increase in the orbital period. Note that, comparing instantaneous values, the change in the period caused by the wind is $2 - 3$ orders of magnitude higher than the change in the period due to the flow of matter.

\subsection{Outflow from the Lagrange point \texorpdfstring{$L_2$}{L2}}\label{model_L2}
\noindent

As stated earlier, the outflow of matter can occur not only from the disc, but also from the outer Lagrange $ L_2 $, which is located behind the less massive component of the system. We use the work of~\citet{Soberman97} to estimate the change in the period and get:
    \begin{equation}\label{P_L2_P}
    \frac{\dot{P}_{L_2}}{P_1} = \frac{ 3 \dot{M}_{L_2}} {M} \left( \frac{M^2}{M_{\rm opt} \, M_{\rm x} } \sqrt{\frac{R_{L2}}{A}} \right)\, ,
    \end{equation}
where $\dot{M}_{L_2}$~($\dot{M}_{L_2} < 0$) --- the rate of matter outflow from the Lagrange point $ L_2 $, $ R_{L_2} $ is the distance from $ L_2 $ to the center of mass of the system~\citep{EmelSaly83}:
    \begin{equation}
    \begin{gathered}
    \frac{R_{L_2}}{A} \approx \frac{1}{1+1/q} + \left(\frac{1/q}{3(1+1/q)}\right)^{1/3} \! \! \! \! +  \frac{1}{3}\left(\frac{1/q}{3(1+1/q)}\right)^{2/3} \! \! \! \! - \\  -\;\frac{1}{9}\left(\frac{1/q}{3(1+1/q)}\right)  + \frac{50}{81}\left(\frac{1/q}{3(1+1/q)}\right)^{4/3}\, .
    \end{gathered}
    \end{equation}
It can be seen that the outflow from the second Lagrange point leads to a decrease in the period.

\section{Results}\label{results}
\noindent

Using formula~\eqref{pb_p}, we can determine the change in the orbital period due to the wind from the disc and accretion of matter by setting the values of the masses of the system components, $M_{\rm x}$ and $M_{\rm opt}$, the average rates outflow of matter and accretion $\left<\dot{M}_{\rm wind}\right>$ and $\left<\dot{M}_{\rm acc}\right>$, disc size before the burst $R_{\rm disc}$, as well as the effective radius of the disc $R_{\rm esc}$, from which the outflow occurs. For the characteristic radii of the problem, it is convenient to introduce the parameters $k_{\rm disc} \equiv R_{\rm disc}/A$ and $k_{\rm esc} \equiv R_{\rm esc}/A$.

If the angular momentum of matter passing through the Lagrange point $ L_1 $ does not change during the formation of the disc~(which is accurate enough for typical values of $ q $ in the LMXB), then the matter forms a ring with the circulization radius $R_{\rm circ}$ rotating  with the Keplerian speed. We assume that the disc in a quiescent state before the outburst does not spread out, and its characteristic radius $ R_{\rm disc} $ remains equal to $R_{\rm circ}$. Then, according to~\citet{Frank02}~(formula 4.20):
    \begin{equation}\label{rcirc}
    R_{\rm disc} = R_{\rm circ} =  \left(1+\frac1q \right)\left[0.500-0.227 \lg \left(\frac1q\right)\right]^4 \, A\,. 
    \end{equation}
Using~\eqref{rcirc}, we find that, for example: $k_{\rm circ} \equiv R_{\rm circ}/A = 0.226,\; 0.307 $ and $0.420$ for $q = 5,\; 7$ and $20$, respectively.

At the beginning of an outburst, due to the fast heating of the matter in the ring and increase in viscosity, a redistribution of the angular momentum begins: a part of the matter with a decreasing specific angular momentum falls along a spiral to the center, another part with a large specific angular momentum moves away, as a result of which the disc expands to its maximum size. In conservative accretion discs without a wind, almost all the angular momentum from the disc is pumped from its outer radius to the orbital momentum of the binary system~\citep[e.g.,][]{IchSus94}.

To estimate the maximum possible effect of the influence of the wind, let us assume that all the wind starts from the outer boundary of the disc, where the matter has the greatest specific angular momentum. The outer boundary of the disc is acted upon by tidal forces from the side of the neighboring component; therefore, the accretion disc does not reach the boundaries of the Roche lobe of the compact object. As, for example, in~\citet{suleimanov_etal2008}, we choose the outer radius of the accretion disc equal to the tidal radius $ R_{\rm tid} $, which is about $ 90 \% $ of the radius of the Roche lobe $ R_{\rm RL} $, defined by the Eggleton's formula~\citep{eggl}:
    \begin{equation}\label{resc}
    R_{\rm esc} = R_{\rm tid} = 0.9 \times R_{\rm RL} = 0.9 \times \frac{0.49q^{2/3}}{0.6q^{2/3} + \ln{(1+q^{1/3})}} \ A \ ,
    \end{equation} 
where $q \equiv M_{\rm x} / M_{\rm opt}$ is the ratio of the masses of the compact (for which the Roche lobe is considered) and optical components of the close binary system. Using~\eqref{resc}, we obtain: $k_{\rm tid} \equiv R_{\rm tid}/A = 0.469,\; 0.520 $, and $0.567$ for $q = 5,\; 7$, and $20$, respectively.

Note that, due to the fact that the formulas~\eqref{rcirc}  and~\eqref{resc} are approximate and are obtained from various considerations, for sufficiently large values of the mass ratio $ q $ (starting from $ q \approx 47.8 $), the circulization radius exceeds the outer radius of the disc, which is unrealistic. Therefore, we impose the following condition on the radius of the disc before an outburst: $R_{\rm disc} \leq R_{\rm esc}$.

Having obtained the estimates~\eqref{rcirc}  and~\eqref{resc} of the characteristic radii of the problem, let us investigate how the effect of changing the period of the binary system depends on the relative power of the wind from the disc. Let us assume that the average accretion rate onto a compact object is $\left<\dot{M}_{\rm acc}\right> = 10^{18}$ g/s, which, by order of magnitude, is the typical accretion rate during an LMXB outburst  and is about one tenth of the critical Eddington accretion rate onto a non-rotating black hole with mass $M_{\rm x} = 10\,M_{\odot}$.
    \begin{figure}[ht!]
    \begin{center}
    \center{\includegraphics[width=\columnwidth]{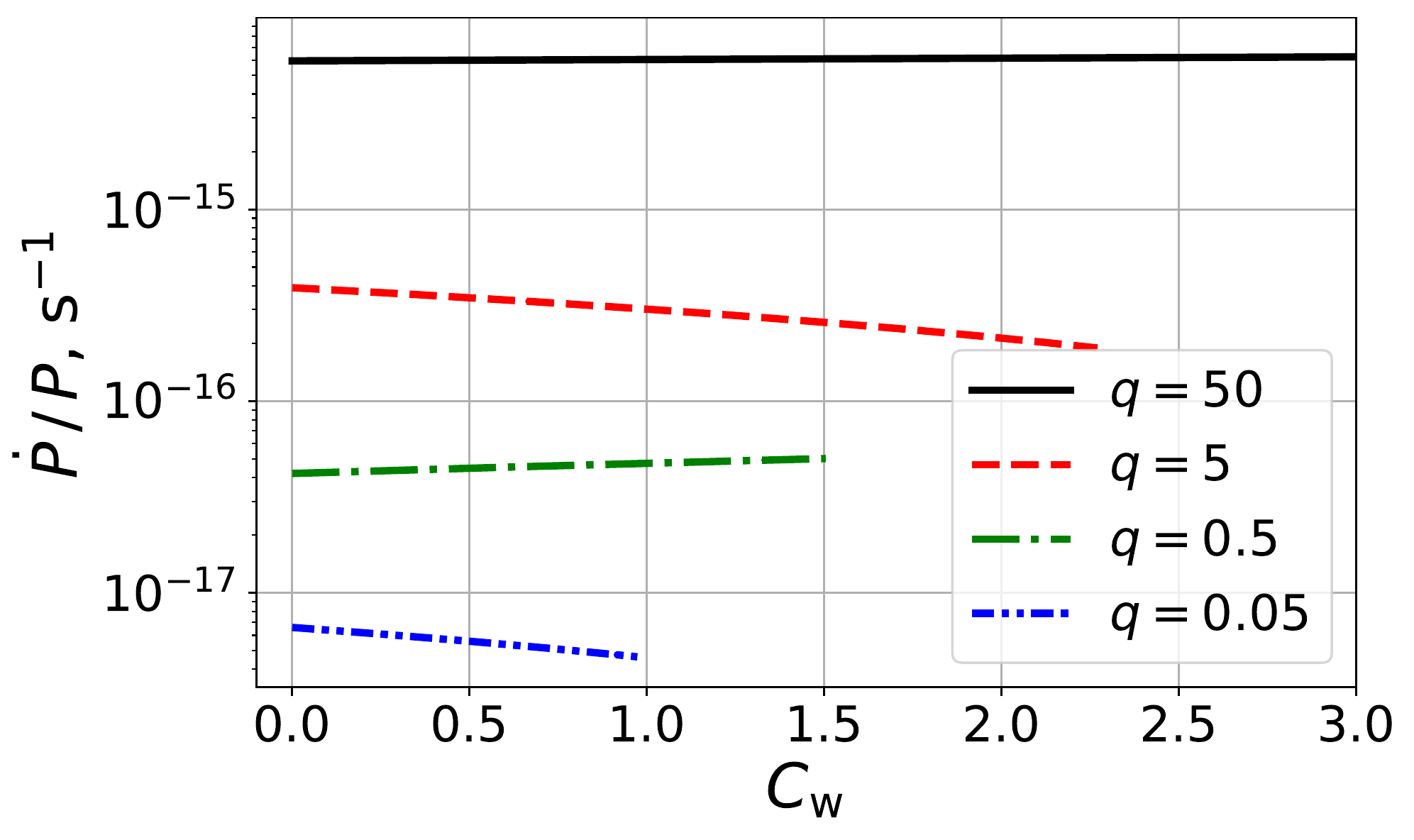}}
    \caption{Relative change in the orbital period of a close binary system versus the ratio of the rates of mass loss due to wind and accretion for different mass ratios if $R_{\rm esc}=R_{\rm tid}$. The average accretion rate $\left<\dot{M}_{\rm acc}\right>$ is $10^{18}$~g/s.}
    \label{p_pacc}
    \end{center}    
    \end{figure}

Let the rate of the mass loss in the wind be proportional to the accretion rate  onto the black hole: $\left|\left<\dot{M}_{\rm wind}\right> \right|= C_{\rm w} \times \left<\dot{M}_{\rm acc}\right>$. Fig.~\ref{p_pacc} shows the rates of the relative change in the orbital period according to~\eqref{pb_p} depending on the ratio $ C_{\rm w} $ for various mass ratios of the components $ q $ for $R_{\rm esc}=R_{\rm tid}$  and $R_{\rm disc}=R_{\rm circ}$. Curves end on the right when $C_{\rm w}=C_{\rm w}^{*}(q)$. However, for $q \gtrsim 47.8$ the value of $C_{\rm w}^{*}$ does not exist, since the approximation formulas for the tidal and circulation radius give a non-physical relation (see the formula \eqref{C_wmax}). Therefore, the curves in the given range of values of $q$~($ q \gtrsim 47.8 $)  are not limited by any maximum $C_{\rm w}$.

Note also that on the intervals $q \lesssim 0.1 $ and $1.1 \lesssim q \lesssim 42.7$ the curves are decreasing functions of $ q $, since for these values of $q$ the following condition is satisfied: $k_{\rm esc}^{1/2} > k_{\rm disc}^{1/2} + 2/3 \, (q^{1/2}/( 1 + q )^{3/2})$. 
    \begin{figure}[ht!]
    \begin{center}
    \center{\includegraphics[width=\columnwidth]{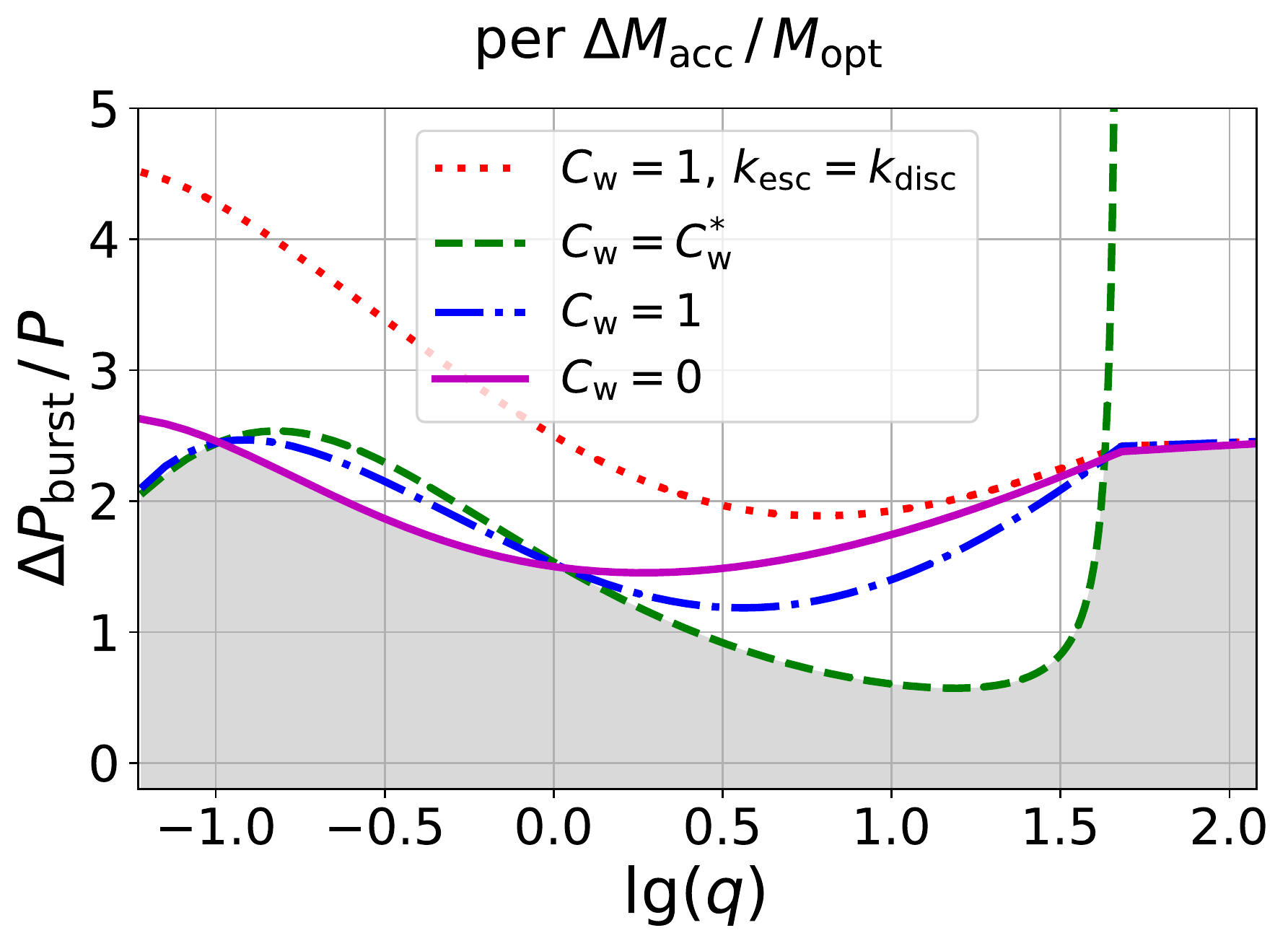}
    }
    \caption{
    Normalized period variation $ (\Delta P_{\rm burst}/P )\,(\Delta M_{\rm acc}/M_{\rm opt})^{-1}$ 
    depending on the logarithm of the component mass ratio $q=M_{\rm x}/M_{\rm opt}$.
    The solid curve is the change in the period if there is no wind from the disc ($C_{\rm w}=0$), the dashed red curve shows the case when the mass loss in the wind is equal to the accreted mass ($C_{\rm w}=1$), but the effective radius of the wind outflow is equal to the characteristic radius of the disc before the outburst; other curves: the wind starts from the tidal radius of the disc  $R_{\rm tid}$, $C_{\rm w} = 1$ and $C_{\rm w} = C_{\rm w}^{*}$. The area below the minimum possible period change is shown in gray.}
    \label{norm_period_change}
    \end{center}
    \end{figure}
    
Figure~\ref{norm_period_change} shows the value of the normalized change in the period $\frac{\Delta P}{P} \,(\frac{\Delta M_{\rm acc}}{M_{\rm opt}})^{-1}$ depending on the mass ratio $ q $. As can be seen from the figure, the presence of the wind for the interval of values $ q $, typical for LMXB, leads to some decrease in the predicted period gain due to the outburst, comparing with the conservative model.

\subsection{Applying the model to real LMXB systems}\label{results_2}
\noindent

\begin{center}
    \begin{figure}
    \center{\includegraphics[width=\columnwidth]{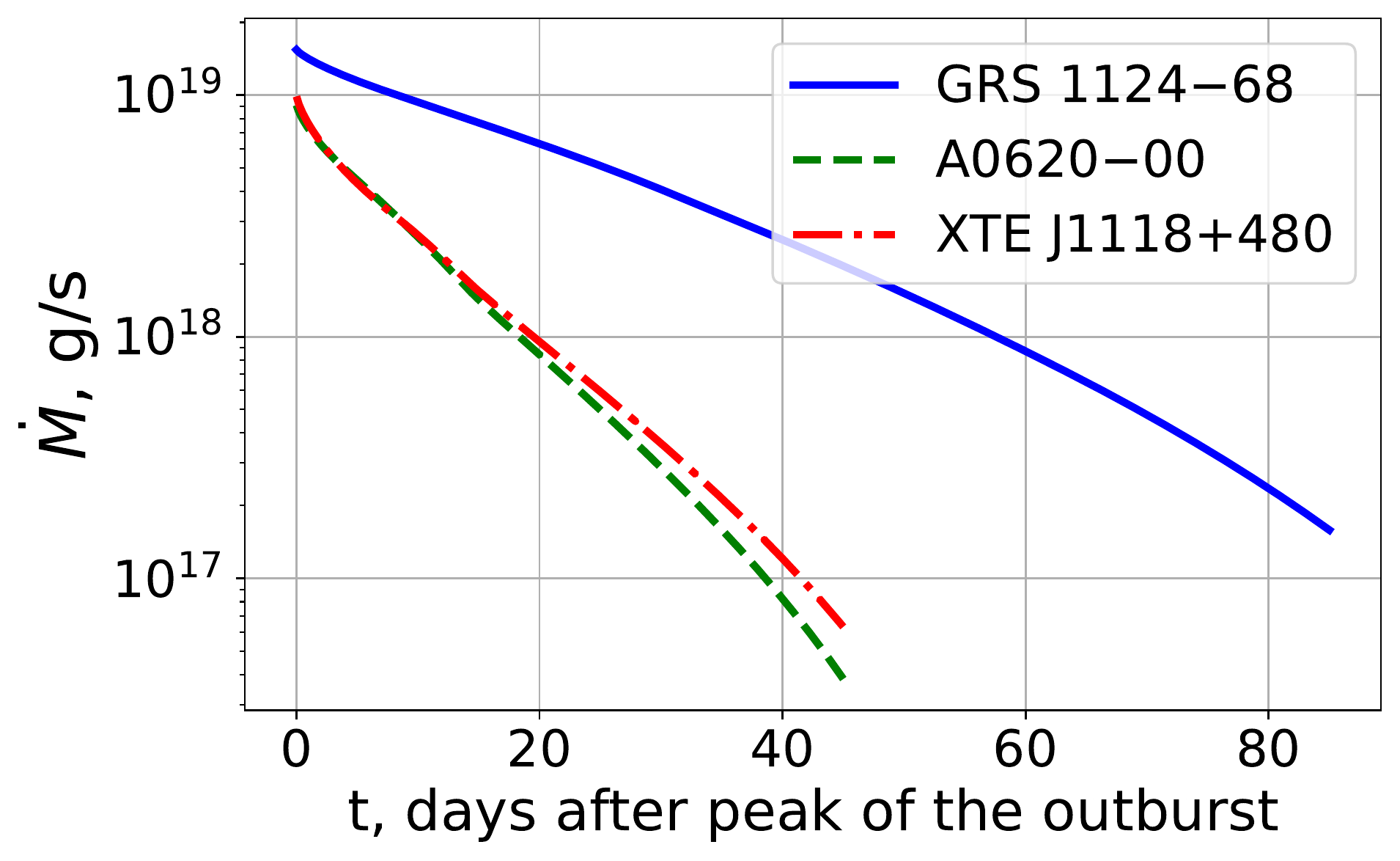}}
    \caption{Simulated outbursts evolution with disc wind for parameters of \ursa{}, \mon{}, and \musc{}. The initial accretion rate equals to the Eddington value. The ratio of the rate of mass loss due to wind and accretion is $C_{\rm w} = 7, \, 8.45, \, 3.87$, respectively. The system parameters are listed in Table~\ref{tab:params_1}.}
    \label{mdot_sys}
    \end{figure}
    \end{center}

Using the code \textsc{freddi}\footnote{\textsc{freddi} can be freely downloaded from the web page \url{http://xray.sai.msu.ru/~malanchev/freddi}.} \citep{Mal_Lip16, lipunova_malanchev2017}, artificial outbursts of \ursa{}, \mon{}, and \musc{}~were simulated, taking into account disc wind~(see Fig.~\ref{mdot_sys}). Model parameters for systems are shown in Table.~\ref{tab:params_1}. Note that the initial accretion rate for all three systems is taken equal to the critical Eddington rate.

The \textsc{freddi} code calculates the viscous evolution of a disc zone that is fully ionized. The code was developed to calculate the light curves of soft X-ray transients with rapid growth and quasi-exponential decay and was modified for this work to take into account the effect of wind from the disc~\citep{Avakyan19}.

    \begin{table*}
    \caption{\label{tab:params_1} Parameters of binary systems \ursa{}, \mon{}, and \musc{} used in simulations.}
    \renewcommand{\arraystretch}{1.7}
    \renewcommand{\tabcolsep}{1.2mm}
    \begin{center}
    \begin{tabular}{c|c|c|c|c}
    \hline
    {\bf Parameters} & {\bf \ursa{}}  & {\bf \mon{}} &  {\bf \musc{}} & {\bf Ref.*} \\
    \hline
    Black Hole Mass, $M_{\rm x}$ & $7.06 M_{\odot}$ &  $ 6.5 M_{\odot}$ & $11.0 M_{\odot}$ & [1], [2], [3]\\
    \hline
    Companion Star Mass, $M_{\rm opt}$ & 
    $ 0.10 M_{\odot}$ & $ 0.26  M_{\odot}$ & $ 0.89 M_{\odot}$ & [1], [2], [3] \\
    \hline
    Orbital period, $P$ &  $ 0.1699$ d  &  $ 0.3230$ d & $ 0.4326$ d& [4], [5], [6]\\
    \hline
    Inclination, $i$ & $74.0^\circ$  &  $ 51.0^\circ$ &  $43.2^\circ$ & [2], [7], [3]\\
    \hline
    Viscosity parameter, $\alpha$ & $0.1$ & $0.1$ & $0.1$ & [8] \\
    \hline
    \end{tabular}
    \end{center}
    * [1]~\citet{Cherep19a}; [2]~\citet{Cherep19b}; [3]~\citet{Wu16}, [4]~GH12; [5]~GH14; [6]~GH17; [7]~\citet{Cantrell10}; [8]~\citet{SHAKURA_SUNAEV}. 
    \\
    {\bf Note.} The peak accretion rate is equal to the Eddington one:
    $\dot{M}_{\rm acc,0} = \dot{M}_{\rm Edd} = 1.4 \times 10^{18} \ (M_{\rm x}/M_{\odot})$ g/s.
    \end{table*}
    
\begin{figure}
    \begin{center}
    \includegraphics[width=\columnwidth]{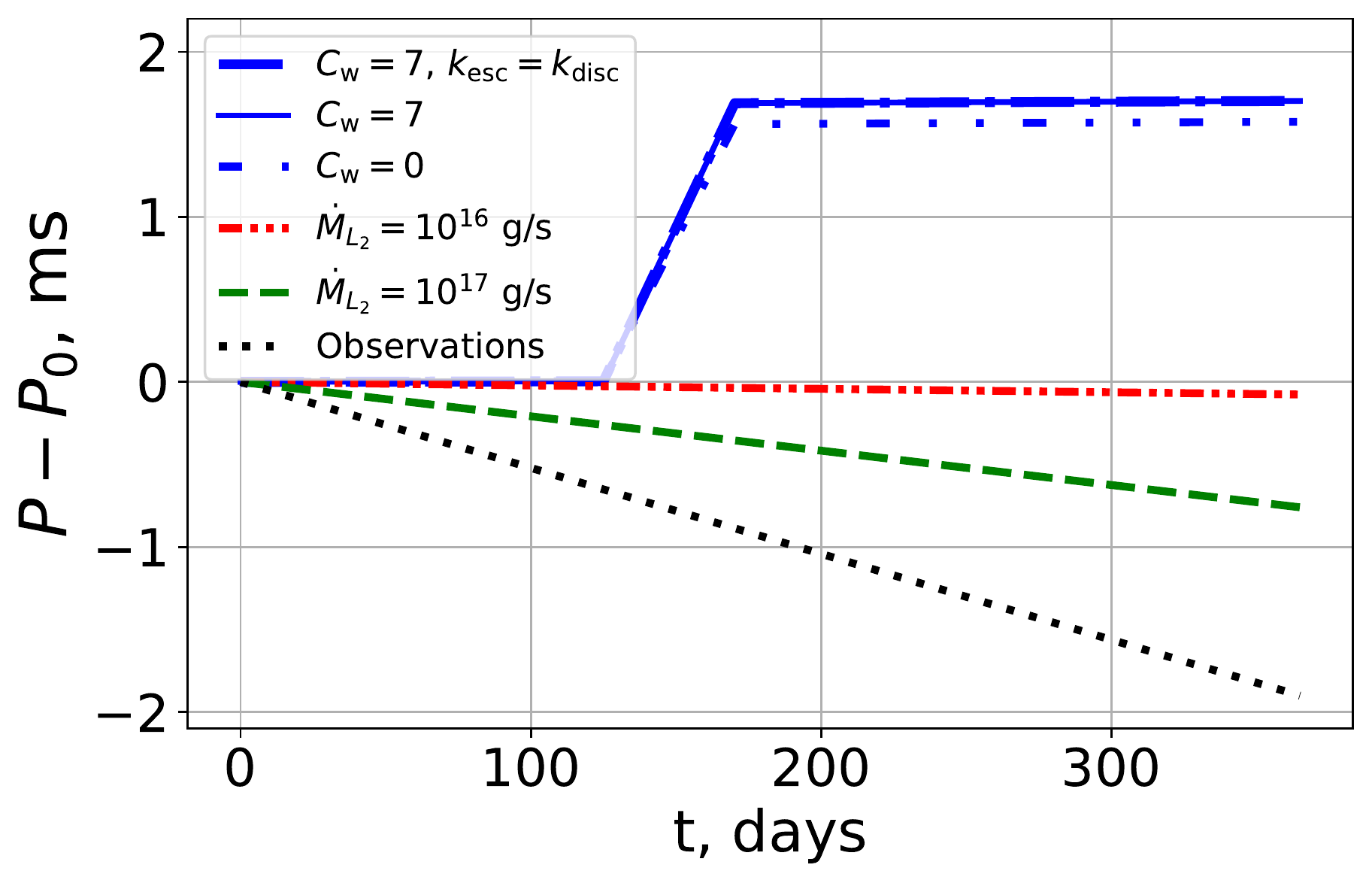}
    \caption{Change in the period for the \ursa{} (within one year) according to the model (blue curves: mass transfer and outburst), observations (black), and constant outflow from the Lagrange point $ L_{2} $ (green and red).}
    \label{periods_XT}
    \end{center}
    \end{figure}

\begin{figure}
    \begin{center}
    \includegraphics[width=\columnwidth]{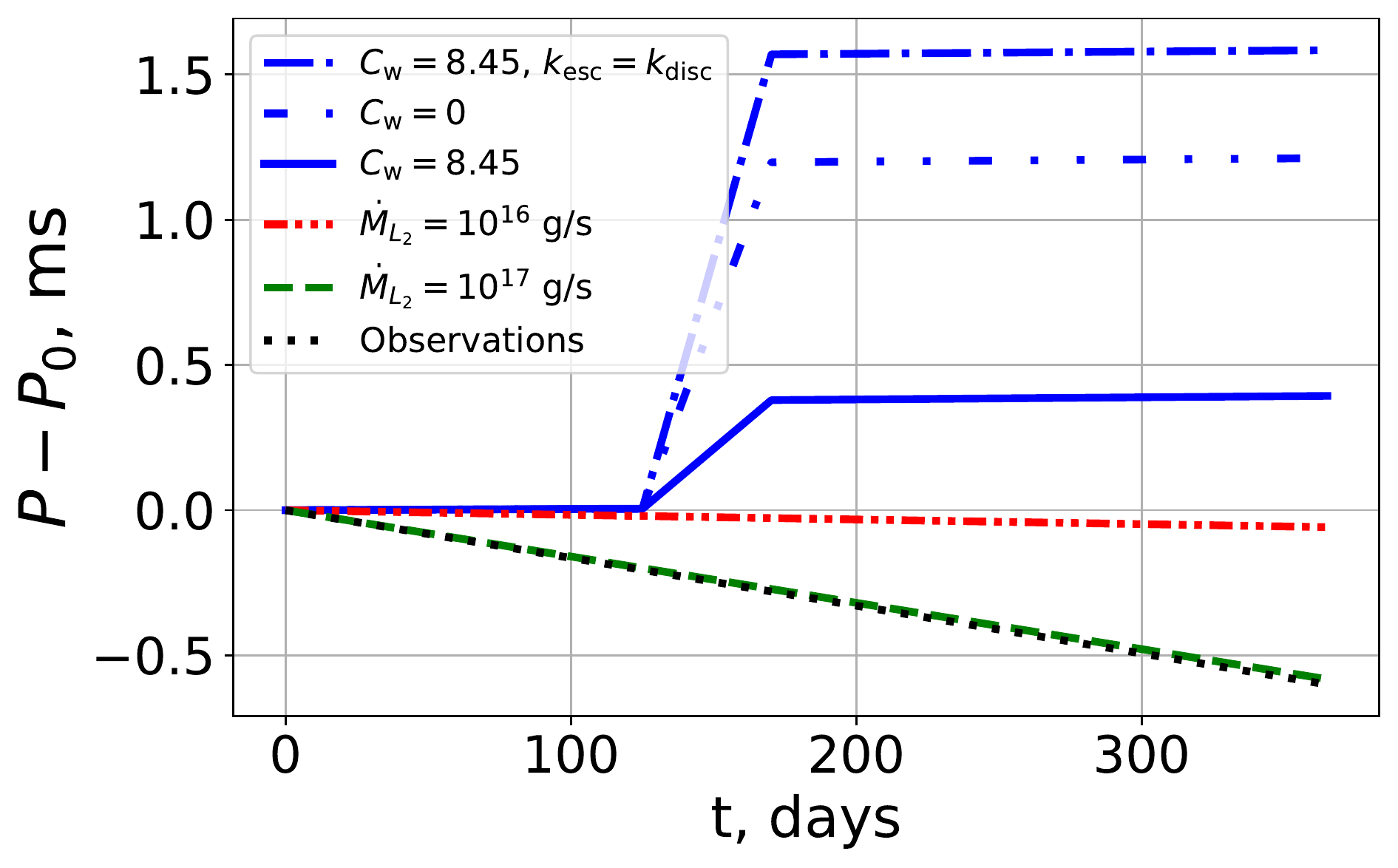}
    \caption{Change in the period for the \mon{} (within one year) according to the model (blue curves: mass transfer and outburst), observations (black), and constant outflow from the Lagrange point $ L_{2} $ (green and red).}
    \label{periods_a0}
    \end{center}
    \end{figure}

\begin{figure}
    \begin{center}    
    \center{\includegraphics[width=\columnwidth]{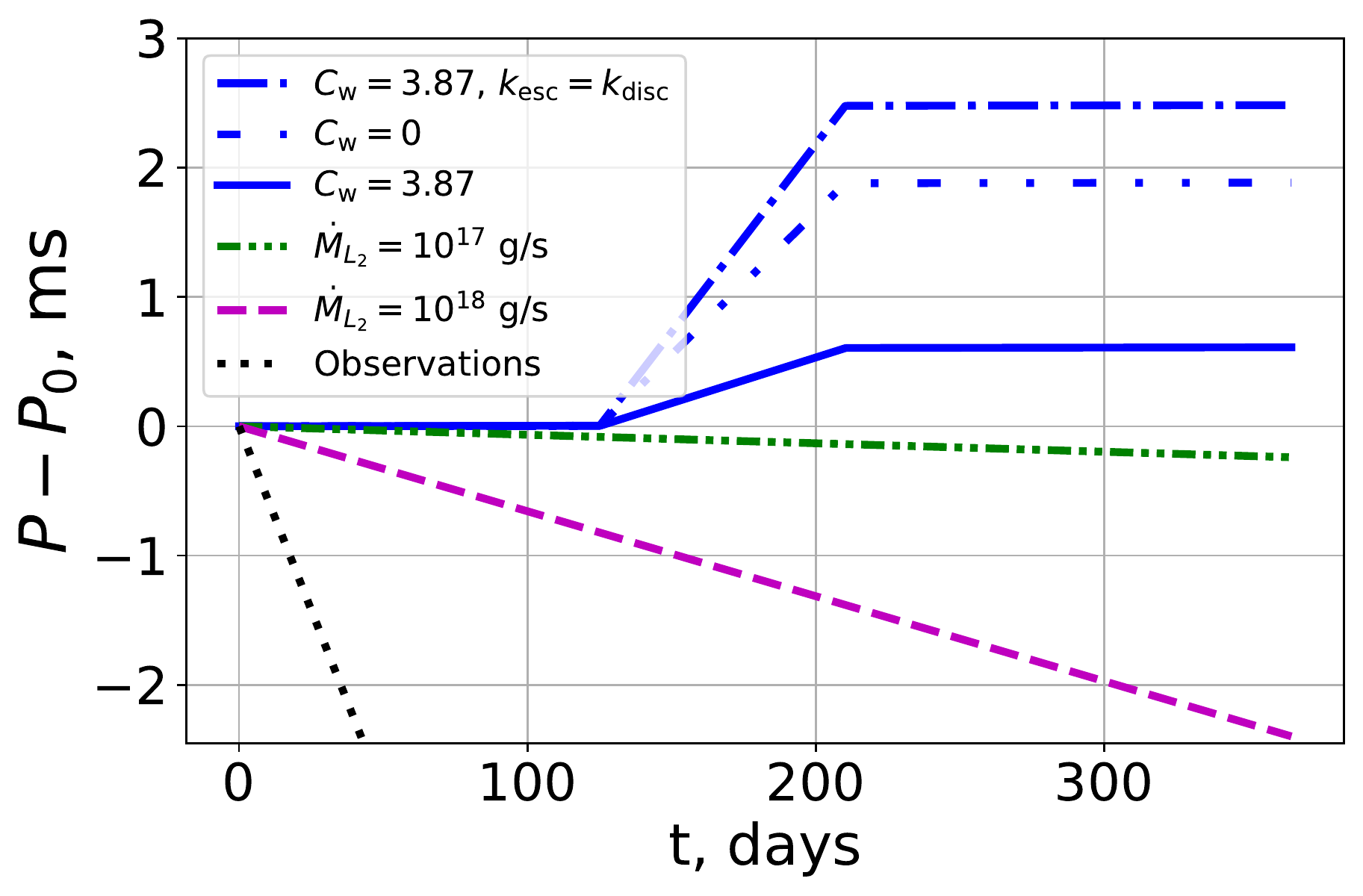}}
    \caption{Change in the period for the \musc{} (within one year) according to the model (blue curves: mass transfer and outburst), observations (black), and constant outflow from the Lagrange point $ L_{2} $ (green and red).}
    \label{periods_grs}
    \end{center}
    \end{figure}

The accretion rates obtained by modeling were used to find an estimate of the change in the orbital period~(Figs.~\ref{periods_XT}, \ref{periods_a0} and \ref{periods_grs}). The ratio of the rate of mass loss in the wind to the accretion rate onto the black hole is chosen as $C_{\rm w} = 7; \, 8.45; \, 3.87$ for \ursa{}, \mon{} and \musc{}, respectively. The maximum possible $C_{\rm w} = C_{\rm w}^{*}(q)$ is taken for \mon{} and \musc{}, appplying $R_{\rm esc} = R_{\rm tid}$ and $R_{\rm disc} = R_{\rm circ}$. Such values lead to the lowest estimate of $ \Delta P $. For parameters of \ursa{}, the definition of $C_{\rm w}^{*}$ is not applicable due to the formal equality of the tidal radius $R_{\rm tid}$ and the circulization radius $R_{\rm circ}$~(i.e., $k_{\rm esc} = k_{\rm disc}$). This happens because the value of the mass ratio of the components in this system $q$ exceeds the boundary value of $ 47.8 $. Therefore, $C_{\rm w}$ was set equal to 7~\citep{Luketic}. The lowest $ \Delta P $ for this system is realized when $C_{\rm w}=0$ (see Fig.~\ref{norm_period_change}).


With these parameters, the average values of the orbital period change during the outburst are about $4.3 \times 10^{-10}$~s/s, $9.6 \times 10^{-11}$~s/s, and $8.2 \times 10^{-11}$~s/s for \ursa{}, \mon{}, and \musc{}, respectively. Such a strong change of the orbital period occurs only during the outburst itself~(about $40-90$ days, we chose $45$ for \ursa{} and \mon{}, and $85$ days for \musc{}). In a steady state (lasting from several to tens of years), the accretion and wind rates in the disc are much lower, and the mechanisms of wind formation can be completely switched off.

Table~\ref{tab:persys}  shows the calculated values of the period variations due to the wind accretion for the \ursa{}, \mon{}, and \musc{} under the assumption of one outburst per  $\Delta T_{\rm q} = 6, \ 30$, and $20$ years, respectively. In other words, the systems, according to the model, are in a state with high accretion rate  for several weeks~(Fig.~\ref{mdot_sys}), when the period is changing significantly, after which a quiescent state begins, lasting $\Delta T_{\rm q}$. Specific values $\Delta T_{\rm q}$ are not chosen by chance. \ursa{} bursts quite often, in 2000 and 2005~\citep{Simon20}, which cannot be said about \mon{} and \musc{}, in which the outbursts have not been observed since 1975 and 1991~\citep{Connors17, Wu16}, respectively. The observations of \ursa{}, used in the works of GH12 and GH14 to determine the decrease in the orbital period, were carried out from 2000 to 2012. In the case of \mon{}, the last measurement of the period was made in 2006~(GH14), and the last outburst occurred in 1975. For \musc{}, the last measurement of the period was in 2012~(GH17), more than twenty years after the outburst in 1991.
    \begin{table*}
    \caption{\label{tab:persys} Orbital changes $\Delta P_{\rm burst}/\Delta T_{\rm q}$ [s/s] due to a `windy' outburst  for \ursa{},  \mon{}, and \musc{}
    as well as the observed and model values of $\dot P$ [s/s] from  GH and CP19.}
    \begin{center}
    \renewcommand{\arraystretch}{1.7}
    \renewcommand{\tabcolsep}{1.2mm}
    \begin{tabular}{c|c|c|c}
    \hline
    {\bf Parameters} & {\bf \ursa{}}  & {\bf \mon{}} &  {\bf \musc{}} \\ 
    \hline
    Our lower estimate & $ 6.6 \times 10^{-11} $ &  $ 3.9. \times 10^{-13} $ & $ 9.5 \times 10^{-13} $ \\
    \hline
    Observations~(GH) & $ -6.0 \times 10^{-11} $ & $ -1.9 \times 10^{-11} $ & $-6.6 \times 10^{-10} $ \\
    \hline
    Model~2~(GH) & $ -5.4 \times 10^{-13} $ &  $ -7.6 \times 10^{-13} $ & $ -8.9 \times 10^{-13} $ \\
    \hline
    Model~1~(GH) & $ -2.7 \times 10^{-11} $ & $ -8.6 \times 10^{-12} $ & $ -3.5 \times 10^{-12} $\\
    \hline
    Gravitational waves (CP19)&  $ -3.0 \times 10^{-13} $ & $ -2.0 \times 10^{-13} $ & $ -4.0 \times 10^{-13} $\\
    \hline
    Magnetic braking (CP19)&  $ -7.8 \times 10^{-12} $ & $ -3.8 \times 10^{-12} $ & $ -2.2 \times 10^{-12} $\\
    \hline
    \end{tabular}
    \end{center}
    \end{table*} 

In addition to our results, Table~\ref{tab:persys} shows long-term values of $\dot P$ from the series of works GH12, GH14, and GH17 (both theoretical and determined from observations, hereinafter referred to as 'GH') and estimates of CP19. In the works of the GH, the model of the  period decrease of a binary by~\citet{Johannsen} is used, which takes into account the magnetic braking and loss of matter due to the evaporation of a black hole. The GH uses two sets of parameters for this model, namely, `realistic' and `extreme' (the maximum possible effect of mass loss and magnetic braking). In Table~\ref{tab:persys}, the designations Model~1 and Model~2 are chosen for the `realistic' and `extreme' set of parameters, respectively. All values are given in order to qualitatively compare the power of various mechanisms that affect the evolution of the orbital periods of binary systems.

Period changes due to outbursts  are hardly noticeable against the background of the observed secular trend in \ursa{}, \mon{} and \musc{} on scales of 6, 30, and 20 years, respectively. Note that the values in the first line of Table~\ref{tab:persys} do not take into account other effects that cause secular changes in the orbital period described above, namely, the outflow from the Lagrange point $ L_2 $ and the transfer of the matter from the optical component to the disc.    
    
Figures~\ref{periods_XT}, \ref{periods_a0}, and \ref{periods_grs} show the period variations in a year when an outburst occurs.
Lines with a `step' show the total effect, namely, the sum of the mechanisms of the matter transfer  (a subtle secular increase in the period for $ \left<\dot M_{\rm tr}\right> = 10^{16}$  g/s)  and the variation during the outburst (the step itself). The effect of the outflow from the point $ L_2 $ is depicted for the two values of the mass loss rate, differing by a factor of $10$. The outflow from the point $ L_2 $ leads to a secular decrease in the period, but the observed rate requires an abnormally high rate of the mass loss. It can be seen that for each of the three systems, the outflow from $ L_2 $ could explain the decrease in the period of the order of the observed one only with very large, unrealistic loss rates through $ L_2 $, especially in the case of \musc{}.

Observations of the orbital period of an X-ray binary  before and after an X-ray nova outburst would be a good evidence in favor of a strong period change during an outburst. According to our model, the orbital periods of  \ursa{}, \mon{}, and \musc{}  can increase by $ \gtrsim 1.6 $, $ \gtrsim 0.4 $, and $ \gtrsim 0.6 $~ms, respectively, however, such observations of the sources have not yet been carried out.
    
\section{Discussion}\label{discus}
\noindent

According to the observations of GH14 and GH17, the three considered LMXB systems with black holes demonstrate the strong period decrease in the quiescent state. The decrease in the period due to the gravitational radiation~(CP19) is $2-3$ orders of magnitude less than the observed rate, and estimates for the magnetic braking are less by an order of magnitude, even with very favorable parameters~(see Table~\ref{tab:persys}). Of the mechanisms discussed above, only the outflow of matter from point $L_2$ with an excessively high rate could explain the observed rates (see Fig.~\ref{periods_XT}, \ref{periods_a0}, and \ref{periods_grs}).

In addition to the mechanisms discussed so far, the following scenario is  possible. The matter ejected by the wind from the accretion disc, or escaping through the $L_2$ point, can form a toroidal disc around the binary system~(a so-called circumbinary disc). Due to the viscosity processes, the  circumbinary disc can drain the moment of tidal forces acting on it, and thereby, reduce the angular momentum of the binary system (therefore decreasing its orbital period). The characteristic viscous time of evolution of a circumbinary disc significantly exceeds the viscous time of the hot disc around a compact object, which allows it to remove the angular momentum on a longer time scale. \citet{Muno_Mauer} and \citet{Wang_XZ} provide observational evidence for the existence of such a disc around \mon{} and \ursa{}. In the work of CP19, based on the results of~\citet{Arty_Lubow}, the influence of such a ring on the period of a close binary system has been studied. According to CP19, the observed decrease in the period of \mon{} and \ursa{} can be explained if the mass of the circumbinary disc is approximately equal to $10^{-9}\,M_{\odot}$. It follows from our simulations that for \mon{} and \ursa{}, during an outburst~(even for the case $C_{\rm w} = 2$), the mass taken with the wind from the disc is, respectively, $ 3.4 \times 10^{-8}\,M_{\odot}$ and $2.4 \times 10^{-8}\,M_{\odot}$. Most likely, not all the matter of the wind settles into a ring around the binary, but even if so, such an estimate may be sufficient to explain the observed decrease in the period for the \mon{} and \ursa{}. However, for \musc{}, the ring around the binary system must be two orders of magnitude heavier, namely, $10^{-7}\,M_{\odot}$.
The required mass can be accumulated after one outburst,  if the wind is very strong~(for $C_{\rm w} = 10$ the mass of matter ejected in the wind is $ 1.1 \times 10^{-7}\,M_{\odot}$). The fact of such a strong change in the \musc{} period makes it even more interesting for a  further study.

\section{Summary}\label{conclus}
\noindent    

An outburst in a low-mass X-ray binary leads to a significant change of its orbital period. We have obtained a general analytical formula for assessing this effect, taking into account the wind from the accretion disc, and have given the quantitative estimates of the magnitude of the period change for three LMXBs.

Observations of various LMXB systems and measurements of the orbital period immediately before and after outbursts can reveal the predicted changes in the orbital period and
provide additional information concerning the rate of matter loss due to wind.

We have also considered the changes in the orbital period due to mechanisms operating in the quiescent state of LMXBs: the transfer of matter from the donor to the disc around the black hole and the outflow of matter from the point $ L_2 $. The latter leads to a secular decrease in the orbital period. However, it also cannot explain the observed rate of the period decrease in the sources considered. Apparently, it is necessary to involve either extremely strong magnetic braking, or the drain of the orbital angular momentum by a ring of matter surrounding a binary system.
    
\section*{Acknowledgements}
\noindent

    The authors are grateful to K.A. Postnov for valuable remarks, the seminar of the Relativistic Astrophysics Department for a productive discussion, and also  to I.~I.~Antokhin for comments. ALA is grateful to Foundation for the Advancement of Theoretical Physics and Mathematics `BASIS' for supporting his work~(grant number 20-2-1-106-1). This research has been supported by the Interdisciplinary Scientific and Educational School of Moscow University ``Fundamental and Applied Space Research".
    The development of the code \textsc{freddi} was supported by the RFBR grant no. 18-502-12025.

\bibliography{main.bib}

\end{document}

%% file: preamble.tex
\usepackage[a4paper,textwidth=18cm,textheight=24cm,top=2.3cm, bottom=2cm, left=1.35cm, right=1.35cm]{geometry}

\usepackage{main}

\makeatletter
\long\def\@makecaption#1#2{%
\vskip\abovecaptionskip
\sbox\@tempboxa{#1. #2}%
\ifdim \wd\@tempboxa >\hsize
#1. #2\par
\else
\global \@minipagefalse
\hb@xt@\hsize{\box\@tempboxa\hfil}%
\fi
\vskip\belowcaptionskip}
\makeatother